\documentclass[aps,preprint]{revtex4}%
\usepackage{amsfonts}
\usepackage{graphicx}
\usepackage{amsmath}
\usepackage{amssymb}%
\providecommand{\U}[1]{\protect \rule{.1in}{.1in}}

\begin{document}
\title[FFLO in a 1D potential]{Resonant enhancement of the FFLO-state in 3D by a one-dimensional optical potential}
\author{Jeroen P. A. Devreese$^{1}$}
\author{Sergei N. Klimin$^{1}$}
\altaffiliation{On leave of absence from: Department of Theoretical Physics, State University
of Moldova, str. A. Mateevici 60, MD-2009 Kishinev, Republic of Moldova.}

\author{Jacques Tempere$^{1,2}$}
\affiliation{$^{1}$TQC, Universiteit Antwerpen, B-2020 Antwerpen, Belgium.}
\affiliation{$^{2}$Lyman Laboratory of Physics, Harvard University, Cambridge, MA 02138, USA.}

\begin{abstract}
We describe an imbalanced superfluid Fermi gas in three dimensions within the
path-integral framework. To allow for the formation of the
Fulde-Ferell-Larkin-Ovchinnikov-state (FFLO-state), a suitable form of the
saddle-point is chosen, in which the pairs have a finite centre-of-mass
momentum. To test the correctness of this path-integral description, the
zero-temperature phase diagram for an imbalanced Fermi gas in three dimensions
is calculated, and compared to recent theoretical results. Subsequently, we
investigate two models that describe the effect of imposing a one-dimensional
optical potential on the 3D imbalanced Fermi gas. We show that this 1D optical
potential can greatly enlarge the stability region of the FFLO-state, relative
to the case of the 3D Fermi gas without 1D periodic modulation. Furthermore it
is show that there exists a direct connection between the centre-of-mass
momentum of the FFLO-pairs and the wavevector of the optical potential. We
propose that this concept can be used experimentally to resonantly enhance the
stability region of the FFLO-state.

\end{abstract}
\date{\today}

\maketitle

\section{Introduction}

The study of superconducting and superfluid systems has recently attracted
wide attention, among other things because of the realization of ultracold
Fermi gasses in optical lattices \cite{Hofstetter, Modugno Inguscio, Koehl,
Chin Ketterle}. These systems can be considered as quantum simulators that can
be used for probing fundamental problems in condensed-matter physics
\cite{Bloch review}, for instance the search for exotic new phases in strongly
magnetized superconductors. Ultracold Fermi gasses offer important advantages
over conventional superconductors, mainly because of their extensive
tunability. In a superconductor, the number of spin-up and spin-down electrons
is equal and the interaction strength is fixed. In ultracold Fermi gasses, one
can not only tune the interaction strength by use of Feshbach resonances
\cite{Timmermans, Andreev Radzihovsky, Esslinger}, but also the population
imbalance can be freely adapted. This experimental freedom has led to the
study of a variety of new phenomena in imbalanced ultracold Fermi gasses
\cite{Zwierlein Ketterle, Shin Ketterle, Partridge Hulet, Hulet en Stoof,
Schunck}. One fundamental question, that is still not settled, concerns the
nature of the ground state of an imbalanced Fermi gas. When population
imbalance between the spin-up and spin-down components is introduced into
these systems, complete pairing is no longer possible. Clogston and
Chandrasekhar suggested that above a critical imbalance, the superfluid system
would undergo a transition into the normal state \cite{Clogston,
Chandrasekhar}. This effect has been observed experimentally by the MIT
\cite{Zwierlein Ketterle} and Rice \cite{Partridge Hulet} groups. However,
their observations were not in exact agreement, and there still exists some
controversy \cite{Chevy} about the exact nature of the phases of the
superfluid system at high levels of imbalance. In 1964 Fulde and Ferell
\cite{Fulde Ferell} and independently Larkin and Ovchinnikov \cite{Larkin
Ovchinnikov} proposed that a superfluid system can accommodate population
imbalance, by making a transition into a state with a finite center-of-mass
momentum (and thus a spatially modulated order parameter). This state is the
so-called Fulde-Ferell-Larkin-Ovchinnikov-state (FFLO-state). Recently, there
has been an ongoing theoretical search for this exotic state \cite{FFLO
theorie}. In 1D, the FFLO-state was predicted to exist in superconducting
systems \cite{Yang}. This work was confirmed numerically
\cite{Heidrich-Meisner 1} and elaborated further through theoretical studies
of the ground state and the phase diagram of a 1D Fermi gas \cite{Hu Liu
Drummond, Orso, Parish}. Furthermore it has been shown that the 1D analogue of
the FFLO-state is stable in a large section of the BCS-BEC crossover phase
diagram \cite{Heidrich-Meisner 2}, compared to the case of a 3D Fermi gas.
Although this 1D FFLO-state has not yet been observed directly, a recent paper
reports the experimental observation of density profiles that agree
quantitatively with theoretical predictions at low temperature \cite{Liao
Rittner}. In three dimensions however, the experimental observation of the
FFLO-state has so far remained elusive. One of the main reasons for this is
that the FFLO-state in three dimensions only occurs in a tiny section of the
BCS-BEC crossover phase diagram \cite{Hu Liu, Sheehy Radzihovsky}. This then
begs the question, is there a way to stabilize the FFLO-state in a 3D Fermi
gas? The purpose of this paper is twofold: first we develop a path-integral
description for a superfluid Fermi gas which can accommodate the FFLO-state,
and second we propose a method to stabilize the FFLO-state through an optical
potential. In two recent papers it was suggested to stabilize the FFLO-state
by the use of a 3D optical lattice \cite{Koponen, Trivedi}. In this paper we
investigate the stabilizing effect of a 1D optical potential in order to
investigate the interplay between the wavevector of the FFLO-state and the
wavevector of the laser which creates the optical potential. In the present
work, the 1D optical potential provides a periodic modulation in one
direction. We emphasize that we do not look at the FFLO-state in a
one-dimensional gas \cite{3D gas}, but in a 3D gas with a superimposed
one-dimensional periodic potential. In section \ref{The free Fermi gas} we
describe the FFLO state in an imbalanced Fermi gas in 3D within the
path-integral framework. As a test for the correctness of this description, we
calculate the zero temperature phase diagram for this system and compare our
findings with recent theoretical results \cite{Hu Liu, Sheehy Radzihovsky}. In
section \ref{Modelling a one-dimensional external potential} we investigate
two models to account for the effect of a one-dimensional optical potential.
We show that the presence of such a potential leads to a substantial increase
of the stability region of the FFLO-state. Finally in section
\ref{Conclusions} we draw conclusions.

\section{Path integral description\label{The free Fermi gas}}

\begin{figure}
[h]
\begin{center}
\includegraphics[
height=9.1951cm,
width=8.9271cm
]%
{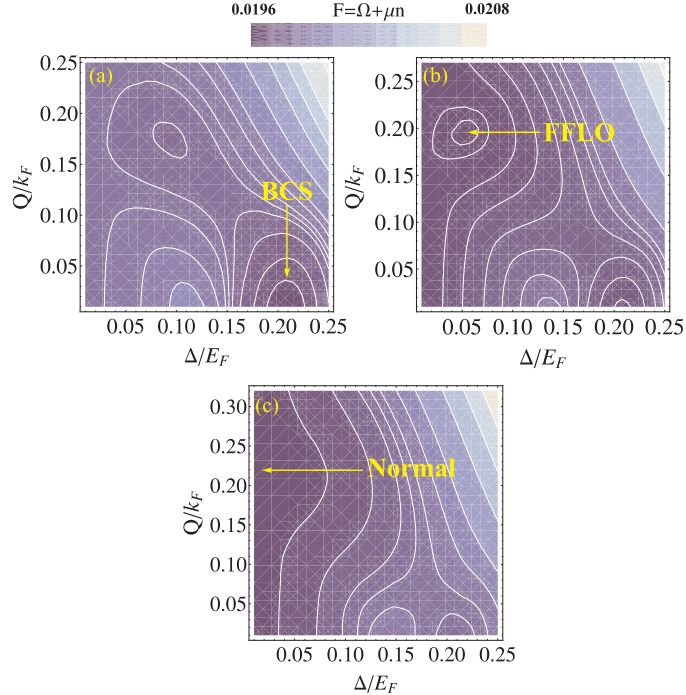}%
\caption{Free energy landscape for an imbalanced Fermi gas in 3D at zero
temperature, for different values of the imbalance chemical potential: (a)
$\zeta/E_{F}=0.141$ (b) $\zeta/E_{F}=0.152$ (c) $\zeta/E_{F}=0.164$. As the
level of imbalance is increased, the competition between the local minima
becomes apparent. Under the right circumstances, the FFLO-state is the ground
state of the system (b). The value of the interaction strength used here is
$\frac{1}{k_{F}a_{S}}=-1$.}%
\label{contours_geenzeta.eps}%
\end{center}
\end{figure}

The partition sum of an imbalanced Fermi gas in 3D can be written as a path
integral over the fermionic fields $\bar{\psi}_{\mathbf{k},\omega_{n},\sigma}$
and $\psi_{\mathbf{k},\omega_{n},\sigma}$:
\begin{equation}%
\begin{array}
[c]{l}%
\mathcal{Z}=%
{\textstyle \int}
\mathcal{D}\bar{\psi}\mathcal{D}\psi \text{ }\exp \left(  -%
{\displaystyle \sum \limits_{\mathbf{k},n}}
{\displaystyle \sum \limits_{\sigma}}
\bar{\psi}_{\mathbf{k},\omega_{n},\sigma}\left(  -i\omega_{n}+\mathbf{k}%
^{2}-\mu_{\sigma}\right)  \psi_{\mathbf{k},\omega_{n},\sigma}\right. \\
\qquad \left.  -\dfrac{g}{\beta L^{3}}%
{\displaystyle \sum \limits_{\mathbf{k},n}}
{\displaystyle \sum \limits_{\mathbf{k}^{\prime},n^{\prime}}}
{\displaystyle \sum \limits_{\mathbf{q},m}}
\bar{\psi}_{(\mathbf{q}/2)+\mathbf{k},\Omega_{m}+\omega_{n},\uparrow}\bar
{\psi}_{(\mathbf{q}/2)-\mathbf{k},\Omega_{m}-\omega_{n},\downarrow}%
\psi_{(\mathbf{q}/2)-\mathbf{k}^{\prime},\Omega_{m}-\omega_{n^{\prime}%
},\downarrow}\psi_{(\mathbf{q}/2)+\mathbf{k}^{\prime},\Omega_{m}%
+\omega_{n^{\prime}},\uparrow}\right)  .
\end{array}
\label{toestandssom begin}%
\end{equation}
Here $\mathbf{k}$ is the wavevector, $\omega_{n}$ are the fermionic Matsubara
frequencies and $\sigma=\uparrow,\downarrow$ denote the two different
hyperfine states. Furthermore, $\beta$ is the inverse temperature given by
$1/k_{B}T$, $L$ is the lateral size of the system, the chemical potential of a
particle with spin $\sigma$ is denoted by $\mu_{\sigma}$ and $g$ is the
renormalized interaction strength. We use units such that $\hbar=2m=E_{F}=1$.
This partition sum (\ref{toestandssom begin}) can be made more tractable by
introducing the Hubbard-Stratonovic transformation, which decouples the
fourth-order interaction term into second order terms by introducing two
auxiliary bosonic fields $\phi_{\mathbf{q},\Omega_{m}}$ and $\bar{\phi
}_{\mathbf{q},\Omega_{m}}$, interpreted as the pair fields. As a first
approximation, only the saddle-point is taken into account in the path
integral over the bosonic fields. To describe the FFLO-state, we propose to
use a saddle point at which the atomic pairs have a finite wavevector
$\mathbf{Q}$:%
\begin{equation}
\phi_{\mathbf{q},\Omega_{m}}=\delta_{\mathbf{q},\mathbf{Q}}\delta_{m,0}%
\sqrt{\beta L^{3}}\Delta~. \label{zadelpunt}%
\end{equation}
By using this particular form of the saddle-point, we choose to describe the
FF-state, which has an order parameter given by a plane wave $\sim
e^{i\mathbf{Qr}}$. It is also possible to describe the LO-state, which is a
superposition of two plane waves with wavevector $\mathbf{Q}$ and
$-\mathbf{Q}$. In this paper, we will not consider the LO-state. For the
remainder of the article, the FF-state is referred to as the FFLO-state. When
$\mathbf{Q}$ is set equal to zero in (\ref{zadelpunt}), the description of the
normal superfluid is recovered \cite{Tempere}. In expression (\ref{zadelpunt}%
), the prefactor $\sqrt{\beta L^{3}}$ ensures that $\Delta$ has units of
energy. Using (\ref{zadelpunt}) the fermionic fields can be integrated out in
expression (\ref{toestandssom begin}) for the partition function, leading to
an effective action $\mathcal{S}_{sp}$, through
\begin{equation}
\mathcal{Z}_{sp}=\exp \left(  \sum_{\mathbf{k},n}\ln \left[  -\det \left(
-\mathbb{G}_{\mathbf{k},n}^{-1}\right)  \right]  +\dfrac{\beta L^{3}}%
{g}\left \vert \Delta \right \vert ^{2}\right)  =\exp \left(  -\mathcal{S}%
_{sp}\right)  ,~ \label{saddle-point partition sum}%
\end{equation}
with $\mathbb{G}_{\mathbf{k},n}^{-1}$ the inverse Nambu propagator which is
given by%
\begin{equation}
-\mathbb{G}_{\mathbf{k},n}^{-1}=\left(
\begin{array}
[c]{cc}%
-i\omega_{n}+(\mathbf{Q}/2+\mathbf{k})^{2}-\mu_{\uparrow} & -\Delta \\
-\Delta^{\ast} & -i\omega_{n}-(\mathbf{Q}/2-\mathbf{k})^{2}+\mu_{\downarrow}%
\end{array}
\right)  .
\end{equation}
Expression (\ref{saddle-point partition sum}) can be simplified further by
performing the sum over the\ Matsubara frequencies. Also, it is useful to
express the results as a function of the total chemical potential $\mu=\left(
\mu_{\uparrow}+\mu_{\downarrow}\right)  /2$ and the imbalance chemical
potential $\zeta=\left(  \mu_{\uparrow}-\mu_{\downarrow}\right)  /2$.
Furthermore, the interaction between particles is modeled with a two-body
contact potential $V\left(  r\right)  =g\delta \left(  r\right)  $. The
renormalized interaction strength $g$ can then be written as follows \cite{De
Melo}:
\begin{equation}
\dfrac{1}{g}=\frac{1}{8\pi \left(  k_{F}a_{s}\right)  }-\sum_{\mathbf{k}}%
\frac{1}{2k^{2}}.
\end{equation}
where $a_{s}$ is the 3D s-wave scattering length. As a final step the
continuum limit is taken, and a thermodynamic potential is associated with the
effective action $\mathcal{S}_{sp}=\beta \Omega_{sp}$. This then results in%
\begin{equation}
\frac{\Omega_{sp}}{L^{3}}=-\int \frac{d\mathbf{k}}{(2\pi)^{3}}\left(  \frac
{1}{\beta}\ln \left[  2\cosh \left(  \beta \zeta_{\mathbf{Q,k}}\right)
+2\cosh \left(  \beta E_{\mathbf{k}}\right)  \right]  -\xi_{\mathbf{Q}%
,\mathbf{k}}-\frac{\left \vert \Delta \right \vert ^{2}}{2k^{2}}\right)
-\frac{\left \vert \Delta \right \vert ^{2}}{8\pi \left(  k_{F}a_{s}\right)  },
\label{thermodynamic potential free fermi gas}%
\end{equation}
where the following notations were introduced
\begin{equation}
\left \{
\begin{array}
[c]{l}%
\xi_{\mathbf{Q},\mathbf{k}}=k^{2}-\left(  \mu-\dfrac{Q^{2}}{4}\right) \\
E_{\mathbf{k}}=\sqrt{\xi_{\mathbf{Q},\mathbf{k}}^{2}+\left \vert \Delta
\right \vert ^{2}}\\
\zeta_{\mathbf{Q,k}}=\zeta+\mathbf{Q\cdot k}%
\end{array}
\right.  . \label{korte notaties vrij fermi gas}%
\end{equation}
The resulting form of the thermodynamic potential
(\ref{thermodynamic potential free fermi gas}) has a similar form as the
original result for the homogeneous 3D Fermi gas derived by Iskin and S\'{a}
de Melo \cite{Iskin de Melo} and coincides with it for $Q\rightarrow0$. In
order to test the correctness of (\ref{thermodynamic potential free fermi gas}%
), the thermodynamic potential is used to calculate the zero temperature phase
diagram of an imbalanced Fermi gas in 3D. This can be done for a fixed number
of particles or for fixed chemical potentials. To transform between these two
descriptions, the number equations, given by%
\begin{align}
-\left.  \frac{\partial \Omega_{sp}}{\partial \mu}\right \vert _{\beta,V}  &
=n=\frac{1}{3\pi^{2}}\label{first number equation}\\
-\left.  \frac{\partial \Omega_{sp}}{\partial \zeta}\right \vert _{\beta,V,\mu}
&  =\delta n \label{second number equation}%
\end{align}
have to be solved. As an example, the phase diagram is calculated for a fixed
density $n$ and for a fixed imbalance chemical potential $\zeta$. To do this,
the first number equation (\ref{first number equation}) is solved, given
values of $\zeta,\Delta$ and $Q$, to determine the chemical potential $\mu$.
The set of values $\left(  \mu,\zeta,\Delta,Q\right)  $ is then substituted in
the free energy $\mathcal{F}=\Omega_{sp}+\mu n$. The minima in the free energy
landscape determine which state is the ground state of the system, for a given
imbalance chemical potential $\zeta$ and a given interaction strength
$1/k_{F}a_{s}$. There are three local minima that can be identified in the
free energy landscape: the BCS-state (spin-balanced superfluid) with
$\Delta \neq0$, $Q=0$, the FFLO-state with $\Delta \neq0$, $Q\neq0~$and the
normal state with $\Delta=0.$ Figure \ref{contours_geenzeta.eps} shows that
the FFLO-state can indeed be the ground state of an imbalanced Fermi gas in
three dimensions (at zero temperature). In this figure, the free energy of the
system is shown, as a function of the bandgap $\Delta$ and the wavevector
$\mathbf{Q}$, relative to the Fermi energy $E_{F}$ and the Fermi wavevector
$k_{F}$ respectively.%

\begin{figure}
[h]
\begin{center}
\includegraphics[
height=6.256cm,
width=8.9249cm
]%
{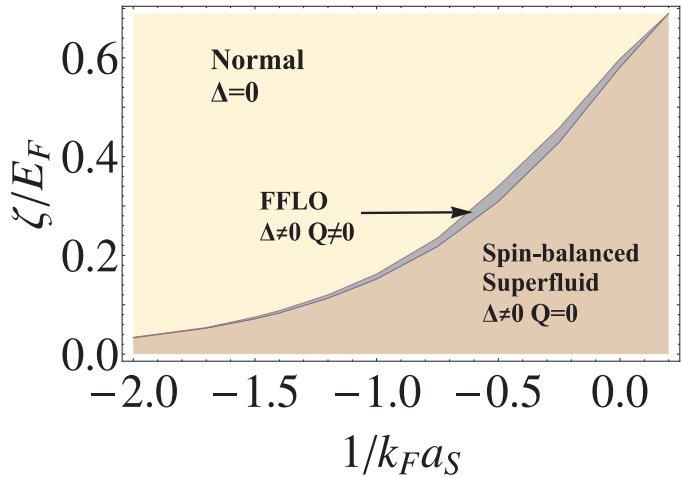}%
\caption{Phase diagram of an imbalanced Fermi gas in 3D at zero temperature,
for fixed density $n$. As the imbalance chemical potential $\zeta$ increases,
the system undergoes a first order transition from a spin-balanced superfluid
(BCS) to the FFLO-state. Above a critical imbalance (which is dependent on the
interaction strength), the FFLO-state continuously goes over into the normal
state.}%
\label{fd_vrijfermigas}%
\end{center}
\end{figure}

For relatively small values of the imbalance chemical potential $\zeta$, the
system is in the BCS ground state (a). When $\zeta$ increases, the system
undergoes a first order transition into the FFLO-state (b). When the imbalance
chemical potential increases further, the system continuously goes over into
the normal state (c). Figure \ref{contours_geenzeta.eps} only shows these
phase-transitions for one specific value of the interaction strength, near the
BCS-limit $\left(  1/k_{F}a_{S}=-1\right)  $. However, since the first number
equation is used to calculate the value of the chemical potential $\mu$, the
description is also valid for the complete BCS-BEC crossover regime. It must
be noted that the mean field approximation breaks down in the unitarity limit.
However, the FFLO-state is expected to form only in the BCS region of the
BCS-BEC crossover, where we expect the mean field description to give
qualitatively correct results. The phase diagram of the system is shown in
figure \ref{fd_vrijfermigas}. This diagram shows that, theoretically, the
FFLO-state can be formed in an imbalanced Fermi gas in 3D, but since it only
occurs on a tiny section of the phase diagram, it may be hard to observe this
state experimentally. Our results coincide with recent theoretical results
\cite{Hu Liu,Sheehy Radzihovsky}.%

\section{Modeling a one-dimensional optical
potential\label{Modelling a one-dimensional external potential}}

As shown in section \ref{The free Fermi gas}, the problem with detecting the
FFLO-state in an imbalanced Fermi gas in 3D is that it exists only in a
relatively small section of the BCS-BEC phase diagram. In this section, we
describe the 3D imbalanced Fermi gas in a 1D optical potential. There are two
main reasons why such a potential can stabilize the FFLO-state. The first
reason is that in an imbalanced Fermi gas in 3D, the FFLO-state can have a
wavevector $\mathbf{Q}$ in an arbitrary direction. This freedom of choice
leads to low-energy bosonic excitations, or Goldstone modes, which render the
FFLO-state unstable. In the presence of a 1D optical potential however, it
will be energetically favorable for the FFLO-state to form in the direction of
the optical potential. This will limit the choice for the wavevector
$\mathbf{Q}$ to just one value, thus suppressing the Goldstone modes, which is
expected to stabilize the FFLO-state. The second reason is that the optical
potential will enhance the 1D modulation of the FFLO-order parameter. We
therefore expect the enhancement to be largest when the wavevector of the
FFLO-state is equal to the wavevector of the optical potential. The present
mean-field treatment does not include the effect of excitations such as the
Goldstone modes, but it does include the energy lowering of the modulated
order parameter due to the optical potential.

In this section we propose two approaches of modeling a 3D imbalanced Fermi
gas in a 1D optical potential. In both approaches, the optical potential is
described by using a modified dispersion relation. In section
\ref{anisotropic masses} we model the optical potential by introducing an
anisotropic effective mass in the direction of the potential \cite{Salomon}
(from here on this is supposed to be the $z$-direction). This approximation is
valid when the Fermi energy of the system lies near the bottom of the lowest
Bloch band, i.e. in the case of low density or a short-wavelength optical
potential. In section \ref{Bloch-dispersion}, we model the optical potential
by treating the full lowest Bloch band in the tight binding approximation
\cite{Menotti Stringari, Modugno, Zwerger, Koetsier}. This model is valid when
the Fermi energy lies in the lowest Bloch band (otherwise more bands have to
be considered), but contrary to the first case, it does not need to lie at the
bottom of the band. In both section \ref{anisotropic masses} and section
\ref{Bloch-dispersion} the optical potential is supposed not to forbid
tunneling, as this would inhibit the formation of the FFLO-state. This implies
that we will treat the imbalanced Fermi gas in 3D in a 1D optical potential as
a three dimensional system with a one-dimensional periodic modulation.

\subsection{Anisotropic effective mass\label{anisotropic masses}}

\begin{figure}
[h]
\begin{center}
\includegraphics[
height=5.7002cm,
width=8.9271cm
]%
{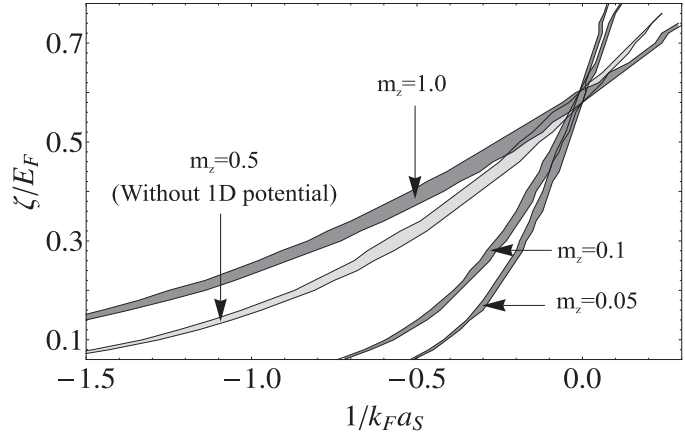}%
\caption{Phase diagrams of 3D imbalanced Fermi gasses in a 1D optical
potential (at temperature zero), where the optical potential was modeled by
altering the effective mass $m_{z}$ (in the z-direction) of the fermionic
particles. Here, the phase diagrams are shown for different effective masses
$m_{z}$. The shaded regions indicate the FFLO phase. These phase diagrams are
shown before rescaling according to (\ref{herschaling}) and hence the
densities differ for the various systems with different effective masses.
After rescaling, all FFLO-regions map onto the case of $m_{z}=1/2$ (indicated
in light gray).}%
\label{mz_ok.eps}%
\end{center}
\end{figure}

In this section, the 1D optical potential is modeled through the use of a
modified effective mass of the fermionic particles in the direction of the
optical potential
\begin{equation}
\varepsilon \left(  k,k_{z}\right)  =k^{2}+\frac{k_{z}^{2}}{2m_{z}},
\label{dispersie anisotroop}%
\end{equation}
where $m_{z}$ is the effective mass of the particles in the $z$-direction.
Here and for the remainder of the paper, $k^{2}=k_{x}^{2}+k_{y}^{2}$ denotes
the magnitude of the in-plane wavevector. This is the wavevector which lies
perpendicular to the laserbeam which creates the 1D optical potential. The
derivation of the thermodynamic potential for this case is analogous to the
derivation in section \ref{The free Fermi gas}. In the present derivation
however, it is assumed that the FFLO-state will form in the $z$-direction
$\mathbf{Q=}\left(  0,0,Q\right)  $, because this is energetically favorable
to other directions, due to the anisotropy introduced through
(\ref{dispersie anisotroop}). The resulting thermodynamic potential, in the
limit for temperature going to zero, is given by%
\begin{equation}
\frac{\Omega_{sp}}{L^{3}}=-\frac{1}{\left(  2\pi \right)  ^{2}}\int
_{0}^{+\infty}dk~k~\int_{-\infty}^{+\infty}dk_{z}\left(  \max \left[
\left \vert \zeta_{\mathbf{k,Q}}\right \vert ,E_{\mathbf{k}}\right]
-\xi_{\mathbf{k}}-\frac{\left \vert \Delta \right \vert ^{2}}{2\left(
k^{2}+\dfrac{k_{z}^{2}}{2m_{z}}\right)  }\right)  -\frac{\left \vert
\Delta \right \vert ^{2}}{8\pi \left(  k_{F}~a_{s}\right)  },
\label{thermodynamic potential anisotropic masses}%
\end{equation}
with the modified notations
\begin{equation}
\left \{
\begin{array}
[c]{l}%
\xi_{\mathbf{Q},\mathbf{k}}=k^{2}+\dfrac{1}{2m_{z}}\left(  k_{z}^{2}%
+\dfrac{Q^{2}}{4}\right)  -\mu \\
E_{\mathbf{k}}=\sqrt{\xi_{\mathbf{Q},\mathbf{k}}^{2}+\left \vert \Delta
\right \vert ^{2}}\\
\zeta_{\mathbf{Q,k}}=\dfrac{1}{2m_{z}}k_{z}~Q-\zeta
\end{array}
\right.  .
\end{equation}
The number equations are still given by (\ref{first number equation}) and
(\ref{second number equation}), but the density $n$ has changed to
\begin{equation}
n=\frac{\sqrt{2m_{z}}}{3\pi^{2}} \label{dichtheid}%
\end{equation}
because of the modified dispersion relation (\ref{dispersie anisotroop}). In
the limit $m_{z}\rightarrow1/2$ the thermodynamic potential
(\ref{thermodynamic potential anisotropic masses}) and the density
(\ref{dichtheid}) converge to the corresponding expressions in the case of an
imbalanced Fermi gas in 3D, described in section \ref{The free Fermi gas}.
Expression (\ref{dichtheid}) implies that when the effective mass $m_{z}$
changes, the density $n$ changes with it. It would be interesting however, to
compare the phase diagrams for Fermi gasses with different effective masses at
equal density. This can in fact be achieved, because we have found that the
thermodynamic potential of the system with an effective mass $m_{z}$ can be
rescaled to the thermodynamic potential of the system with effective mass
$m_{z}=1/2$ (\ref{thermodynamic potential free fermi gas}), using the
following scaling relation%
\begin{equation}
\Omega_{sp}\left(  \mu,\zeta,\Delta,Q,m_{z},\frac{1}{a_{s}}\right)
=\sqrt{2m_{z}}\Omega_{sp}\left(  \mu,\zeta,\Delta,\dfrac{Q}{\sqrt{2m_{z}}%
},\frac{1}{2},\dfrac{1}{a_{s}}\dfrac{1}{\sqrt{2m_{z}}}\right)  .
\label{herschaling}%
\end{equation}
From a theoretical point of view, this rescaling property is time-saving for
calculations and gives a deeper insight into the role of the effective mass
$m_{z}$. The main advantage is however, that all physical properties can be
studied at the same density. This property relates to experiment, because when
an external potential is turned on, the effective mass is altered, but the
average density will remain the same. The effect of changing $m_{z}$ on the
BCS-BEC crossover phase diagram is shown in figure \ref{mz_ok.eps}. This
figure shows the FFLO phase boundaries of the imbalanced Fermi gas in 3D for
different values of the effective mass $m_{z}$, before rescaling according to
(\ref{herschaling}) (and hence at different densities).\
There is a tilting of the FFLO-region about a fixed point at unitarity.
Furthermore, there is an increase in the width of the FFLO-region (relative to
the abscissa) as the effective mass increases. By using the scaling-relation
(\ref{herschaling}), the phase diagrams in figure \ref{mz_ok.eps} can be
rescaled to equal density. After rescaling, the phase diagrams for the
different effective masses maps onto the phase diagram of the imbalanced Fermi
gas with isotropic effective mass ($m_{z}=1/2$), described in section
\ref{The free Fermi gas}. From this we conclude that the FFLO-state is not
fundamentally influenced by an optical potential in which the Fermi energy
lies near the bottom of the first Bloch-band. This can be explained by the
fact that no fundamental anisotropy is introduced into the system by altering
the effective mass, because independent of the effective mass, the system can
be scaled back to the case of the imbalanced Fermi gas where the effective
mass equals 1/2.

\subsection{Bloch-dispersion\label{Bloch-dispersion}}

\begin{figure}
[h]
\begin{center}
\includegraphics[
height=6.0934cm,
width=8.9271cm
]%
{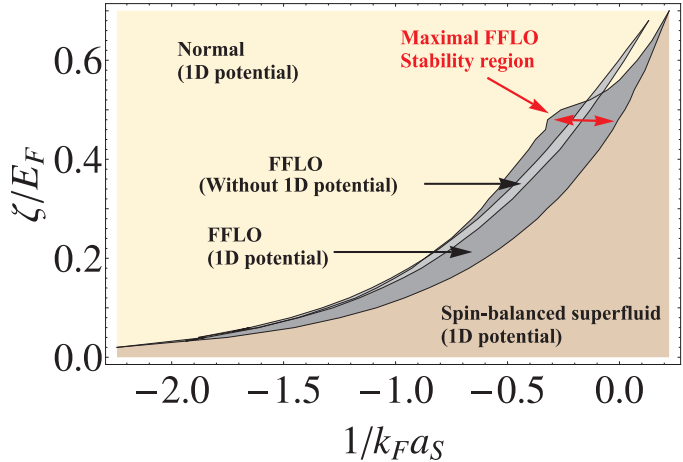}%
\caption{Comparison between the phase diagram of an imbalanced Fermi gas in 3D
and an imbalanced Fermi gas in 3D subjected to a 1D optical potential (with
$\delta=0.5$ and $Q_{L}=1.2$), modeled in the tight-binding approximation,
using the first Bloch band. Both phase diagrams are at temperature zero. The
stabilizing effect of the optical potential enlarges the FFLO region by a
factor 3 to 6 when compared to the case of the imbalanced Fermi gas without
optical potential. When $\zeta/E_{F}\approx0.48$, the width of the FFLO-region
is maximal (for this specific choice of $\delta$ and $Q_{L}$). This resonant
enhancement of the FFLO-region occurs when the wavevector of the FFLO-pairs is
equal to the wavevector of the optical potential.}%
\label{fd_mettekst.eps}%
\end{center}
\end{figure}

In section \ref{anisotropic masses} it was shown that a more fundamental
anisotropy is needed, in order for the optical potential to have an effect on
the FFLO-state. In this section, we model a 1D optical potential in the
tight-binding approximation, using the first Bloch-band. For this purpose, the
quadratic dispersion in the $z$-direction is replaced by a periodic dispersion
\cite{Menotti Stringari, Modugno, Zwerger, Koetsier}
\begin{equation}
\varepsilon \left(  k,k_{z}\right)  =k^{2}+\delta \left[  1-\cos \left(
\frac{\pi k_{z}}{Q_{L}}\right)  \right]  ~. \label{Bloch-dispersie}%
\end{equation}
Here $Q_{L}$ is the wavevector of the optical potential and $\delta$ is a
prefactor with units of energy, given by \cite{Koetsier}%
\begin{equation}
\delta=8\left(  \frac{V_{0}^{3}E_{R}}{\pi^{2}}\right)  ^{\frac{1}{4}}%
\exp \left(  -2\sqrt{\frac{V_{0}}{E_{R}}}\right)  \label{tight binding delta}%
\end{equation}
with $V_{0}$ the depth of the potential and $E_{R}$ the recoil energy given by
$E_{R}=2\pi^{2}\hbar^{2}/m\lambda^{2}$, with $\lambda$ the wavelength of the
optical potential and $m$ the mass of the fermionic particles. In the limit
for small $k_{z}$ expression (\ref{Bloch-dispersie}) simplifies to
(\ref{dispersie anisotroop}), with%
\begin{equation}
m_{z}=\frac{Q_{L}^{2}}{\delta \pi^{2}}. \label{relatie mx Ql delta}%
\end{equation}
Given the new dispersion (\ref{Bloch-dispersie}), the thermodynamic potential
for this system can be calculated. The result is%
\begin{align}
\frac{\Omega_{sp}}{L^{3}}  &  =-\frac{1}{\left(  2\pi \right)  ^{2}}\int
_{0}^{+\infty}dk~k~\int_{-Q_{L}}^{+Q_{L}}dk_{z}\nonumber \\
&  \times \left(  \max \left[  \left \vert \zeta_{k,Q}\right \vert ,E_{\mathbf{k}%
}\right]  -\xi_{\mathbf{k}}-\frac{\Delta^{2}}{2\left \{  k^{2}+\delta \left[
1-\cos \left(  \frac{\pi k_{z}}{Q_{L}}\right)  \right]  \right \}  }\right)
-\frac{\Delta^{2}}{8\pi \left(  k_{F}~a_{s}\right)  }
\label{thermodynamic potential Bloch dispersion}%
\end{align}
with the following notations%
\begin{equation}
\left \{
\begin{array}
[c]{l}%
\xi_{\mathbf{Q},\mathbf{k}}=k^{2}+\delta \left[  1-\cos \left(  \frac{\pi}%
{2}\frac{Q}{Q_{L}}\right)  \cos \left(  \frac{\pi k_{z}}{Q_{L}}\right)
\right]  -\mu \\
E_{\mathbf{k}}=\sqrt{\left \{  k^{2}+\delta \left[  1-\cos \left(  \frac{\pi}%
{2}\frac{Q}{Q_{L}}\right)  \cos \left(  \frac{\pi k_{z}}{Q_{L}}\right)
\right]  -\mu \right \}  ^{2}+\Delta^{2}}\\
\zeta_{\mathbf{Q,k}}=\zeta-\delta \sin \left(  \frac{\pi}{2}\frac{Q}{Q_{L}%
}\right)  \sin \left(  \frac{\pi k_{z}}{Q_{L}}\right)
\end{array}
\right.
\end{equation}
It can easily be shown that expression
(\ref{thermodynamic potential Bloch dispersion}) is equal to the corresponding
thermodynamic potential (\ref{thermodynamic potential anisotropic masses}) of
the anisotropic effective mass case, in the limit $Q_{L}\rightarrow \infty$,
$\delta \rightarrow \infty$ with $m_{z}$ held constant, according to
(\ref{relatie mx Ql delta}). The two number equations again are given by
(\ref{first number equation}) and (\ref{second number equation}) and the
density $n$ can be calculated using the general expression%
\begin{equation}
n=2\int \frac{d\mathbf{k}}{\left(  2\pi \right)  ^{3}}\Theta \left(  1-\left \{
k^{2}+\delta \left[  1-\cos \left(  \frac{\pi k_{z}}{Q_{L}}\right)  \right]
\right \}  \right)
\end{equation}
which yields%
\begin{equation}
n=\left \{
\begin{array}
[c]{c}%
\dfrac{Q_{L}}{2\pi^{2}}\left(  1-\delta \right)  ~\left(  1\geq2\delta \right)
\\
\dfrac{Q_{L}}{2\pi^{3}}\left[  \left(  1-\delta \right)  \arccos \left(
\dfrac{\delta-1}{\delta}\right)  +\delta \sqrt{1-\left(  \dfrac{\delta
-1}{\delta}\right)  ^{2}}\right]  ~\left(  1<2\delta \right)
\end{array}
\right.  ~. \label{n}%
\end{equation}
Here it must be noted that our derivation is only exact if $E_{F}<2\delta$,
because otherwise more than one Bloch band has to be considered. As in the
previous sections, the phase diagram for this system can be constructed by
studying the local minima of the free energy. Figure \ref{fd_mettekst.eps}
shows a comparison between the phase diagram of an imbalanced Fermi gas in 3D
and the phase diagram of an imbalanced Fermi gas in 3D subject to a 1D optical
potential, modeled in the tight-binding approximation using the first Bloch band.%

\begin{figure}
[h]
\begin{center}
\includegraphics[
height=6.2384cm,
width=8.9284cm
]%
{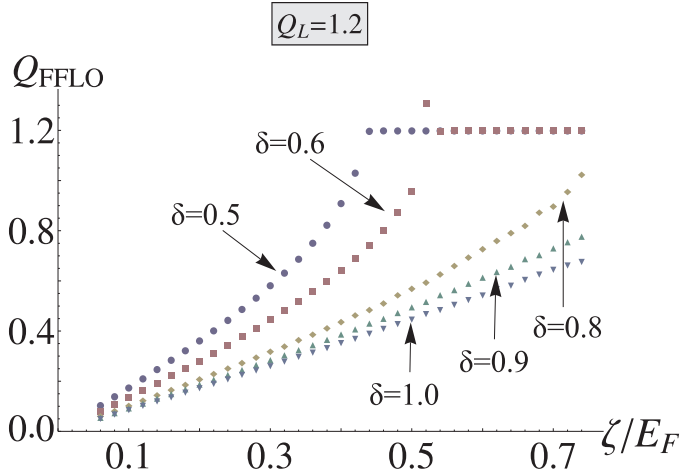}%
\caption{The wavevector of the FFLO-state $Q_{FFLO}$, as a function of the
imbalance chemical potential $\zeta$ (relative to the Fermi energy $E_{F}$).
The value of $Q_{FFLO}$ increases with increasing imbalance, until it reaches
the value of the wavevector of the optical potential $Q_{L}$ (in this case
$Q_{L}=1.2$), which is the saturation point. The single overshoot point for
$\delta=0.6$ is probably a numerical inaccuracy.}%
\label{deltas_ql12.eps}%
\end{center}
\end{figure}

\begin{figure}
[h]
\begin{center}
\includegraphics[
height=18.4312cm,
width=8.9261cm
]%
{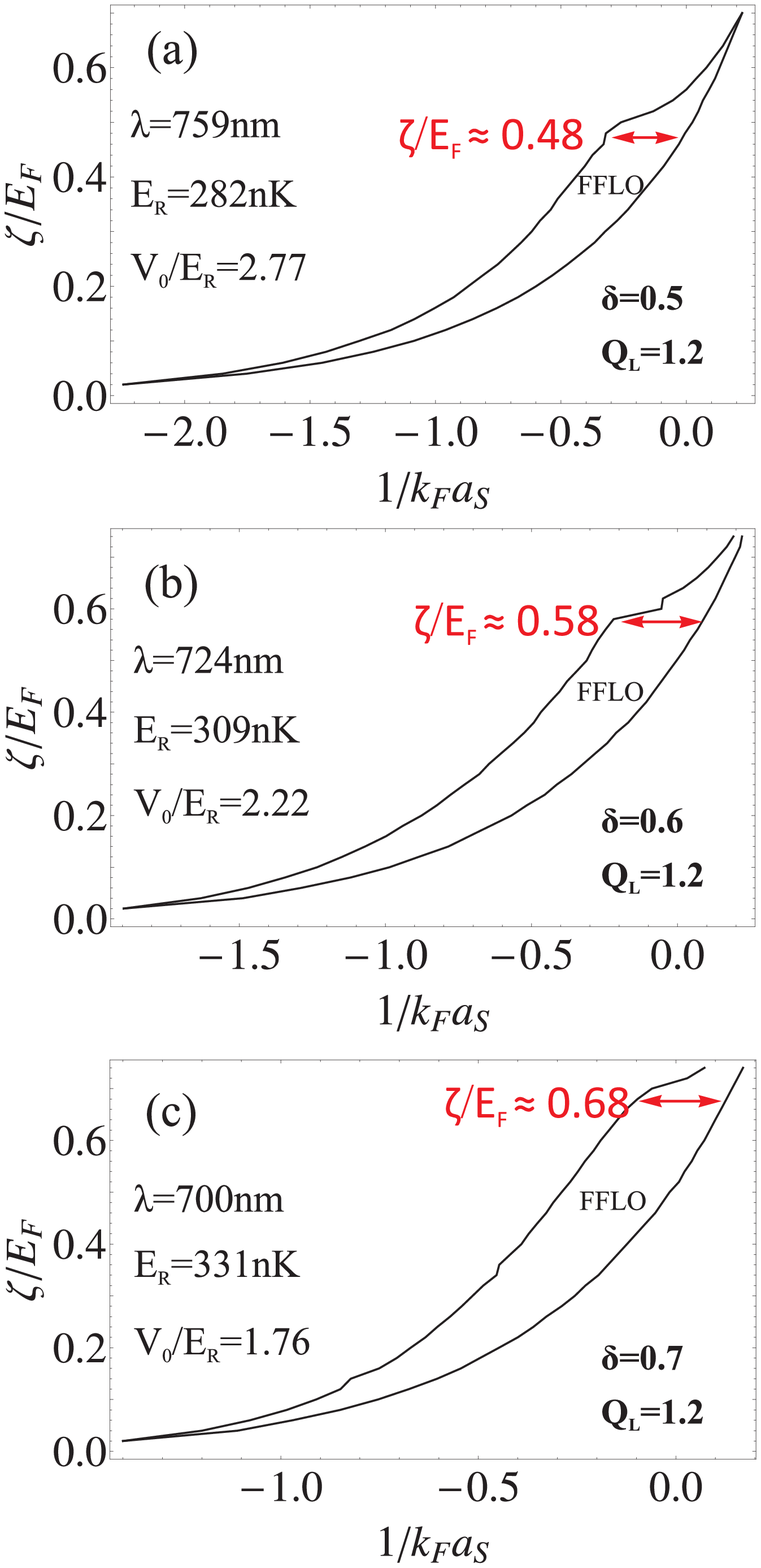}%
\caption{Several phase diagrams of an imbalanced superfluid Fermi gas in 3D
subjected to a 1D optical potential, for increasing values of $\delta$. The
value of the imbalance chemical potential $\zeta$ at which the maximal
FFLO-region occurs, roughly scales linearly with the value of $\delta$. For
each level of imbalance, an optimal FFLO-region can be found, by tuning the
wavelength of the optical potential.}%
\label{delta050607.eps}%
\end{center}
\end{figure}

This figure shows that the FFLO-region is enlarged by a factor 3 to 6, due to
the stabilizing effect of the 1D optical potential. Figure
\ref{fd_mettekst.eps} further shows that at $\zeta/E_{F}\approx0.48$ a
transition point occurs, where the FFLO-region reaches a maximum width
(relative to the abscissa), and narrows quickly for larger values of $\zeta$.
This effect finds its origin in the magnitude of the wavevector of the
FFLO-pairs $Q_{FFLO}$. When the imbalance chemical potential $\zeta$
increases, $Q_{FFLO}$ increases likewise to accommodate for the widening gap
between the Fermi surfaces of the two spin-species. At a certain level of
imbalance (in the case of figure \ref{fd_mettekst.eps} at $\zeta/E_{F}%
\approx0.48$) $Q_{FFLO}$ equals the wavevector of the optical potential
$Q_{L}$. At this point, the FFLO-state is optimally enhanced, because the
spatial modulation of the FFLO-state is equal to the spatial modulation of the
optical potential. This results in a maximal width of the FFLO-region. When
$\zeta$ increases further, $Q_{FFLO}$ retains the constant value $Q_{L}$, and
is not able to grow any further. This effect is shown in figure
\ref{deltas_ql12.eps}.%
Hence, we can conclude that, although the imbalance has increased further, the
optical potential forces the FFLO-state into a state where the form of the
FFLO-order parameter matches the form of the optical potential. This results
in a narrowing of the FFLO-region, because the wavevector of the FFLO-state is
not sufficiently large anymore to bridge the gap between the Fermi-surfaces of
the spin-up and spin-down particles. The value of the transition point, where
$Q_{FFLO}$ becomes equal to $Q_{L}$, rougly increases linearly with the value
of $\delta$, as shown in figure \ref{delta050607.eps}.
Qualitatively, this is because the rate of change of the FFLO-wavevector
$Q_{FFLO}$ with increasing imbalance chemical potential $\zeta$, decreases
when $\delta$ becomes larger. This means that $\zeta$ has to be larger
(compared to the case of lower $\delta$) for $Q_{FFLO}$ to reach the limiting
wavevector of the optical potential $Q_{L}$. The advantage of this resonant
enhancement of the FFLO-state is that, for a given level of imbalance, an
optimal stability region for the FFLO-state can be created, simply by tuning
the wavelength of the 1D optical potential. It should be noted that, when
considering an imbalance chemical potential $\zeta$ smaller than $0.48$,
$\delta$ has to become smaller than $0.5$ and more bands have to be taken into
account in order for our description to be exact. We do not treat this case in
the present paper.

To obtain a more direct link with experimental parameter values, we convert
our units back to SI units for the three situations depicted in figure
\ref{delta050607.eps}. A possible choice of atoms which we considered is
$^{40}K$ atoms in a one-dimensional harmonic trap, with a density of
$10^{13}~cm^{-3}$. For instance, the case with $\delta=0.6$ and $Q_{L}=1.2$
corresponds to the case of an optical potential with wavelength equal to
$724~nm$, a recoil energy of $309~nK$ and an optical potential depth of
$V_{0}/E_{R}\approx2.22 $. The numerical values for these experimental
parameters in the case of $\delta=0.5$ and of $\delta=0.7$ are depicted in
figure \ref{delta050607.eps} panels (a) and (c) respectively. For the
illustrative cases of figure \ref{delta050607.eps} we use $Q_{L}=1.2$, but
theoretically, any choice of $Q_{L}$ was possible because we found that the
wavevector of the optical potential $Q_{L}$ acts as a scaling parameter,
according to the following scaling relation:%
\begin{equation}
\Omega_{sp}\left(  \mu,\zeta,\Delta,Q,\delta,\alpha Q_{L},\frac{1}{k_{F}a_{S}%
}\right)  =\alpha~\Omega_{sp}\left(  \mu,\zeta,\Delta,\frac{1}{\alpha}%
Q,\delta,Q_{L},\frac{1}{\alpha}\frac{1}{k_{F}a_{S}}\right)  ~.
\end{equation}
In principle this means that we can vary $Q_{L}$ from zero to infinity.
However, there exist some limitations on this parameter. First there is a
lower limit for $Q_{L}$ because below a certain value of $Q_{L}$ no value of
the optical potential depth $V_{0}$ can satisfy equation
(\ref{tight binding delta}), given values for $\delta$ and for the recoil
energy $E_{R}$. Second, when the depth of the optical potential becomes too
large, particles will be confined in the direction of the optical potential,
thus inhibiting the formation of FFLO-states. This sets an upper limit for the
ratio of $V_{0}/E_{R}$ and subsequently for the value of $Q_{L}$.

During the course of our work, Loh and Trivedi published their results on the
LO-state in a 3D\ cubic lattice \cite{Trivedi}. They found that in a 3D cubic
lattice, the LO-state was more stable than the FF-state. Since we already find
a substantial increase in the FF-state using a 1D optical potential, we expect
that the effect on the LO-state will be similar or larger. It would be
interesting to apply our 1D-potential scheme also to the LO-case.

\section{Conclusions\label{Conclusions}}

We have described the FFLO-state in an imbalanced Fermi gas in 3D within the
path-integral framework, by choosing a suitable saddle-point at which the
atomic pairs have a finite centre-of-mass momentum. As a platform to address
the case of a 3D imbalanced Fermi gas in a 1D optical potential and to
validate our path-integral description, we rederived the zero-temperature
phase diagram for an imbalanced Fermi gas in 3D. For this case, our results
coincide with recent theoretical results. As a proposal to stabilize the
FFLO-state we have studied an imbalanced 3D Fermi gas in a 1D optical
potential. This potential was modeled in two different ways. For the first
model, where we considered anisotropic effective masses, we have found that
this model is a rescaling of the case of a 3D imbalanced Fermi gas with
isotropic effective mass. In the second model, we described the effect of the
1D optical potential using the first Bloch band in the tight-binding
approximation. In this case we have found a substantial increase in the
stability region of the FFLO-state, as compared to the case of the 3D Fermi
gas without the 1D optical potential. Related results were recently found in
the case of a 3D cubic optical lattice \cite{Koponen, Trivedi}. The advantage
of our 1D optical potential scheme, compared to a 3D cubic optical lattice, is
that it allows to find an optimal stability configuration for the FFLO-state
with a given level of imbalance, by tuning the wavelength of the optical
potential. This resonant enhancement of the FFLO-region occurs when the
wavevector of the FFLO-pairs is equal to the wavevector of the optical
potential. This tunability makes a 1D optical potential a suitable
experimental configuration for the stabilization of the FFLO-state. We
therefore propose that this concept can facilitate the experimental
observation of the FFLO-state in an imbalanced Fermi gas in 3D.

\begin{acknowledgments}
Acknowledgments -- The authors would like to thank Vladimir Gladilin and Fons
Brosens for fruitful discussions. This work was supported by FWO-V projects
G.0356.06, G.0370.09N, G.0180.09N, G.0365.08.
\end{acknowledgments}

\end{document}